\documentclass[fleqn,usenatbib]{mnras}

\usepackage{mathptmx}
\usepackage[T1]{fontenc}
\usepackage{ae,aecompl}

\usepackage{graphicx}	% Including figure files
\usepackage{amsmath}	% Advanced maths commands
\usepackage{amssymb}	% Extra maths symbols
\usepackage{natbib}
\title[Radio detections of southern ultracool dwarfs]{Radio detections of southern ultracool dwarfs }

\author[C. Lynch et al.]{
C. Lynch,$^{1,2}$\thanks{E-mail: clynch@physics.usyd.edu.au}
T. Murphy,$^{1,2}$
V. Ravi,$^{4}$ 
G. Hobbs,$^{3}$ 
K. Lo,$^{1,5}$
C. Ward$^{1}$ 
\\
$^1$ Sydney Institute for Astronomy, School of Physics, The University of Sydney, NSW 2006, Australia\\
$^2$ ARC Centre of Excellence for All-sky Astrophysics (CAASTRO)\\
$^3$ Australia Telescope National Facility, CSIRO Astronomy and Space Science, PO Box 76, Epping, NSW 1710, Australia\\
$^4$ Cahill Centre for Astronomy and Astrophysics, MC 249-17, California Institute of Technology, Pasadena, CA 91125, USA\\
$^5$ University College London Genetics Institute, University College London, London WC1E 6BT, United Kingdom
}

\date{Accepted 2016 January 6.  Received 2016 January 5; in original form 2015 September 7}

\pubyear{2016}

\begin{document}
\label{firstpage}
\pagerange{\pageref{firstpage}--\pageref{lastpage}}
\maketitle

\begin{abstract}
We report the results of a volume-limited survey using the Australia Telescope Compact Array to search for transient and quiescent radio emission from 15 southern hemisphere ultracool dwarfs. We detect radio emission from 2\uppercase{mass}W J0004348--404405 increasing the number of radio loud ultracool dwarfs to 22. We also observe radio emission from 2\uppercase{mass} J10481463--3956062 and 2\uppercase{mass}I J0339352--352544, two sources with previous radio detections. The radio emission from the three detected sources shows no variability or flare emission. Modelling this quiescent emission we find that it is consistent with optically thin gyrosynchrotron emission from a magnetosphere with an emitting region radius of (1--2)$R_*$, magnetic field inclination 20$^{\circ}$--80$^{\circ}$, field strength $\sim$10 -- 200 G, and power-law electron density $\sim$10$^4$ -- 10$^8$ cm$^{-3}$.  Additionally, we place upper limits on four ultracool dwarfs with no previous radio observations. This increases the number of ultracool dwarfs studied at radio frequencies to 222. Analysing general trends of the radio emission for this sample of 15 sources, we find that the radio activity increases for later spectral types and more rapidly rotating objects. Furthermore, comparing the ratio of the radio to X-ray luminosities for these sources, we find 2\uppercase{mass} J10481463--3956062 and 2\uppercase{mass}I J0339352--352544 violate the G\"udel-Benz  relation by more than two orders of magnitude. 

\end{abstract}


%%%%%%%%%%%%%%%%%%%%%%%%%%%%%%%%%%%%%%%%%%%%%%%%%%

%%%%%%%%%%%%%%%%% BODY OF PAPER %%%%%%%%%%%%%%%%%%

\section{Introduction}\label{sec:intro}

Surveys of chromospheric H$\alpha$ and coronal  X-ray emission from low mass stars show a steady decline in magnetic activity strength beginning in late-type M dwarfs \citep[e.g.][]{Neuhauser:1999, Gizis:2000, West:2004, Williams:2014, Schmidt:2015}. The strength of activity in these two wavebands is frequently characterised by the ratio of the luminosity in the H$\alpha$/X-ray waveband to the bolometric luminosity \citep{Hawley:1996}. The reduction in activity strength is thought to be associated with a decrease in plasma heating through the dissipation of magnetic fields \citep{Mohanty:2002}. However, recent atmospheric modelling of late-type objects indicates that it is not unreasonable to expect observable H$\alpha$ emission for these objects. The rarefied upper parts of the stellar atmospheres are found to be capable of magnetically coupling despite having a low levels of ionisation \citep{Rodriguez-Barrera:2015}. The decline in the magnetic activity strength traced by H$\alpha$ and X-ray emission does not imply a drop in the fraction of active cool stars over later spectral types. The number of active systems, as indicated by H$\alpha$ emission, is observed to increase across later spectral types and peaks between M9 and L0 objects \citep{Schmidt:2015}. Additionally neither the magnetic field strength or filling factor for late type objects is thought to decrease. In fact the detection of both quiescent and flaring non-thermal radio emission from some of the lowest mass stars and brown dwarfs \citep{Berger:2001, Berger:2002, Berger:2006, Berger:2009, Burgasser:2005, Phan-Bao:2007, Osten:2006a, Mclean:2012}, collectively called ultracool dwarfs, confirms that at least some of these objects are still capable of generating strong magnetic fields. 

Most radio loud ultracool dwarfs have a quiescent component, and in some cases, this component is found to vary with the rotation of the star \citep[e.g.][]{Mclean:2011}. There is still some debate over the nature of the quiescent component where both depolarised electron cyclotron maser (ECM) \citep{Hallinan:2007}  and gyrosynchrotron emission from a non-thermal population of electrons \citep{Berger:2002, Burgasser:2005, Osten:2006b} are proposed sources for this emission.  Furthermore, some radio loud ultracool dwarfs are observed to have strong radio flares that can be periodic. The ECM mechanism is generally accepted to be the source of the pulsed emission since it can account for the high brightness temperature, directivity, and circular polarisation of this emission \citep{Hallinan:2006}. These radio flares are sometimes associated with periodic variations in the optical band \citep{Berger:2009,  Williams:2015, Hallinan:2015}.  Recent simultaneous radio and optical observations of a late-type M dwarf showed that the observed modulation at both wavelengths could be accounted for by a propagating electron beam, powered by auroral currents, striking the stellar atmosphere  \citep{Hallinan:2015}.  This results suggests that aurorae may be ubiquitous signatures of large-scale magnetospheres.

\begin{table*}
	\centering
	\caption{Properties of the Survey Sources} \label{table:target-prop}
	\begin{tabular}{lccrrrrrrl}
		\hline
		 2\uppercase{mass} Number & R.A.  & Decl. &  Spectral & Distance  & $v\sin(i)$   & $L_{bol}$ & $L_x/L_{bol}$ & $L_{H\alpha}/L_{bol}$ &  Reference\textsuperscript{a} \\
 		& & & Type & (pc) & (km s$^{-1}$) & ($L_{\odot}$) & \\
 		\hline
		10481258--1120082 & 10 48 12.8 & -11 20 18.9 & M7.0 & 4.5 & 3.0 & -3.16 & -4.43 & -4.63 & 1,6  \\ 
		14563831--2809473 & 14 56 38.1 & -28 09 53.3 & M7.0 & 7.0 & 8.0 & -3.29 & -4.00 & -4.02 & 1, 5, 7\\ 
		11554286--2224586 & 11 55 42.7 & -22 24 59.6 & M7.5 & 9.7 & 33.0 & -3.30 &-4.40& -4.58 & 1, 6, 12\\ 
		%& & & & & & & -3.5 (f) & &   \\ 
		10481463--3956062 & 10 48 13.5 & -39 56 17.0 & M8.0 & 4.0 & 18.0 & -3.39 & -5.00 & -5.15 &  1, 5, 8\\  
		00244419--2708242 & 00 24 44.1& -27 08 19.7 & M8.5 & 7.71 & 9.0 & -3.25 & --  & -4.62  &  11  \\ 
		0339352--352544 & 03 39 35.5 & -35 25 40.8 & M9.0 & 5.0 & 26.0 & -3.79 & -3.70 & -5.30 & 1, 2 \\ 
		%& & & & & & &  (f) & &   \\ 
		03341218--4953322 & 03 34 13.3 & -49 53 28.6 & M9.0 & 8.20 & --  & -- & -- & $<$5.32 & 11 \\
		0853362--032932 & 08 53 35.9 & -03 29 33.5 & M9.0 & 9.0 & 13.5 & -3.49 & -3.70 & -3.93 & 3 \\ 
		%& & & & & & & -2.5 (f) & & \\
		1507476--162738 & 15 07 47.6 & -16 27 44.9 & L5.0 & 7.3 & 32.0 & -4.23 & $<$-4.50& -8.18 &1, 4 \\ 
		08354256--0819237 & 08 35 42.3 & -08 19 21.7 & L5.0 & 9.0 & 23.0 & -4.60 & -- & -7.42& 1, 4, 5\\
		0004348-404405 & 00 04 35.4 & -40 44 21.8 & L5.0 & 10.0 & 42.0 & -4.67 & -- & -7.42 & 1 \\
		17502484--0016151 & 17 50 24.6 & +00 16 13.7  & L5.5 & 8.0 & -- & -- & -- & -- & 1  \\
		0340094--672405 & 03 40 09.3 & -67 24 08.7 & L8.0 & 9.90 & -- &  -- & -- & -- & 10\\
		02550357--4700509 & 02 55 04.0 & -47 00 54.9 & L8.0 & 4.97 & 67 & -4.80 & $<$-4.70 & $<$-8.28& 9, 12 \\
		02572581--3105523 & 02 57 26.1 & -31 05 50.0 & L8.0 & 9.6 0& -- & -4.82 & --  & -- &1\\
		\hline
		\multicolumn{10}{l}{\textsuperscript{a}\footnotesize{References: (1) \citet{Antonova:2013}; (2) \citet{Berger:2001}; (3) \citet{Berger:2002}; (4) \citet{Berger:2006}; (5) \citet{Burgasser:2005};}}\\
		\multicolumn{10}{l}{\footnotesize{(6) \citet{Mclean:2012}; (7)\citet{Osten:2009}; (8) \citet{Ravi:2011}; (9) \citet{Reid:2008}; (10) \citet{Reiners:2008b}; }}\\
		\multicolumn{10}{l}{\footnotesize{(11) \citet{Reiners:2010}; (12) \citet{Williams:2014}}}

	\end{tabular}
\end{table*}

Radio surveys of ultracool dwarfs have found that about 9\% of these system are radio luminous, with 21 currently known radio loud ultracool dwarfs \citep{Berger:2006, Mclean:2012, Antonova:2013, Route:2013, Kao:2015}. Correlations between the presence of transient or quiescent radio emission and other dwarf properties such as rotation and tracers of magnetic activity at other wavelengths (X-ray and H$\alpha$) are not well established. In fact, the radio luminosity of some detected systems is far in excess of the well-known G\"udel-Benz  relation \citep[][GB]{Gudel:1993a}, an empirically-derived ratio between radio and X-ray luminosity that applies to magnetically active stars over a wide range of spectral types. The deviation from this relation observed in some ultracool dwarfs suggests that the chromospheric evaporation model usually applied to flare stars  \citep{Neupert:1968, Machado:1980,Allred:2006} may not apply to these objects. Furthermore, little is known about the geometry or strength of the magnetic fields in ultracool dwarfs as well as the mechanism that populates the magnetospheres with non-thermal electrons. 

To address these issues, we carried out a volume limited survey of a sample of 15 late-type M and L dwarfs located in the southern hemisphere using the Australia Telescope Compact Array (ATCA). The Compact Array Broadband Backend \citep[CABB;][]{Wilson:2011} allows for a bandwidth of 2 GHz per polarisation in each of two independently tuneable intermediate frequency (IF) bands. These wideband capabilities of ATCA easily provide detailed information about how observed radio pulses and quiescent emission vary in time and frequency. Such a characterisation is required if we want to constrain the magnetospheric parameters and geometry of ultracool dwarfs (section \ref{sec:prop-emission}).  Additionally, to understand general trends of radio emission from ultracool dwarfs with regards to their other physical properties, our observations are augmented with values for projected rotational velocities ($v\sin(i)$), H$\alpha$ and X-ray luminosities from the literature (section \ref{sec:general-trends}). 

\section{Observations and Data Reduction}

The sample of 15 ultracool dwarfs were selected form the all-sky-volume-limited compilations of late-M \citep{Reiners:2009b} and L \citep{Reid:2008} dwarfs. From these two catalogs we selected sources with distances $<$10 pc and located in the Southern Hemisphere. This selection of sources consists of 8 M dwarfs and 7 L dwarfs, with spectral types ranging from M7.0 to L8.0. Additionally, this selection of sources includes 3 known radio loud ultracool dwarfs: 2\uppercase{mass} J1456-2809 \citep{Burgasser:2005, Osten:2009}, 2\uppercase{mass} J10481463--3956062 \citep{Burgasser:2005, Ravi:2011}, and 2\uppercase{mass}I J0339352--352544 \citep{Berger:2001}. Further details about the selected survey targets are given in Table~\ref{table:target-prop}.

\begin{table*}
	\centering
	\caption{Log of Observations} \label{table:obs-log}
	\begin{tabular}{lcclc}
		\hline
		2\uppercase{mass} Number & Observation Date &  ATCA Configuration\textsuperscript{a} & Primary Calibrator & Secondary Calibrator \\
		\hline
		10481258--1120082 & 17 Apr 2010 06:41 -- 17 Apr 2010 16:39 &  6A & PKS B0823-500 & 1045-188   \\ 
		14563831--2809473 &  12 Apr 2010 09:32 -- 12 Apr 2010 21:59   & 6A &PKS B1934-638 &1519-273 \\ 
		11554286--2224586 & 19 Apr 2010 05:49 --  19 Apr 2010 16:22 & 6A &PKS B0823-500 &1143-245 \\ 
		10481463--3956062 &  18 Apr 2010 05:12 --  18 Apr 2010 15:43 &6A  &PKS B0823-500 &1104--445   \\ 
		00244419--2708242 & 24 Apr 2010 18:18 -- 25 Apr 2010 06:11 & 6A  &PKS B1934-638 & 2357-318   \\ 
		0339352--352544 & 30 May 2010 18:55 --  31 May 2010 07:08 & 6C &PKS B1934-638 &0405-331 \\
		03341218--4953322 & 29 May 2010 18:43 -- 30 May 2010 06:30 & 6C & PKS B1934-638&0302-623   \\
		0853362--032932 &30 Apr 2010 04:12 -- 30 Apr 2010 13:36 & 6C  &PKS B1934-638& 0906+015    \\ 
		1507476--162738 & 14 Apr 2010 10:13 -- 14 Apr 2010 21:05  & 6A  & PKS B1934-638&1504-166 \\ 
		08354256--0819237 & 21 Apr 2010 03:16 -- 21 Apr 2010 14:01 & 6A &PKS B1934-638&  0859-140\\
		0004348--404405 &25 Apr 2010 17:47 -- 26 Apr 2010 05:31  & 6A &PKS B1934-638&0022-423\\
		217502484--0016151 & 13 Apr 2010 13:35 -- 13 Apr 2010 22:44 & 6A &PKS B1934-638& 1741-038\\
		0340094--672405 & 22 Apr 2010 22:03 -- 23 Apr 2010 10:00 & 6A &PKS B1934-638& 0252-712  \\
		02550357--4700509 & 27 Apr 2010 17:17 -- 28 Apr 2010 05:20 & 6C &PKS B0823-500 & 0252-549   \\
		02572581--3105523 &  28 May 2010 18:11 -- 29 May 2010 06:02 & 6C&PKS B1934-638 &0237-233 \\
		\hline
		\multicolumn{5}{l}{\textsuperscript{a}\footnotesize{The labels refer to different variants of the antenna spacings; these are defined at:}}\\
		\multicolumn{5}{l}{\footnotesize{ \url{http://www.narrabri.atnf.csiro.au/operations/array_configurations/configurations.html}.}}
	\end{tabular}
\end{table*}

The observations of the 15 ultracool dwarfs were carried out with ATCA in its fully extended 6km configurations during the time period between April and June 2010. During our observations the two IF bands were centred on 5.5 and 9.0 GHz simultaneously. A log of these observations is given in Table~\ref{table:obs-log}. 

The visibility data were reduced using the standard routines in the \textsc{miriad} environment \citep{Sault:1995}. The flux scale and bandpass response were determined from the ATCA primary calibrators, either PKS B1934-638 or PKS B0823-500.   Observations of a bright, compact secondary calibrator was used to calibrate the complex gains and leakage between the orthogonal linear feeds in each antenna. Observations of the secondary calibrators (listed in Table \ref{table:obs-log}) were carried out every 20 minutes, for 1.3 minutes. The flare emission observed from some ultracool dwarfs can be confused with low level radio frequency interference (RFI) peaks. However, the flares from ultracool dwarfs are not strongly linearly polarised.  So to avoid confusing RFI with a source flare, we first identified and flagged RFI in the Stokes Q and U polarisations and then extend these flags to the other polarisations. To carry out this flagging scheme we used the \textsc{miriad} flagging tools \textsc{pgflag} and \textsc{blflag}. 

After calibration, the visibility data for each source was inverted and cleaned using the \textsc{miriad} tasks \textsc{invert}, \textsc{clean}, and \textsc{restor}. Bright sources  located in the same field as the target source were identified and removed. This process involved using the \textsc{clean} components for each of the bright sources, while masking the location of the target ultracool dwarf,  and subtracting the source from the visibility data using the \textsc{miriad} task  \textsc{uvmodel}. The phase centre of the resulting visibility data was shifted to the location of the target ultracool dwarf using the \textsc{miriad} task \textsc{uvedit} and then inverted and cleaned in the standard fashion.  

To search for radio emission from each of the sources, we made images for both the 5.5 and 9.0 GHz frequency bands where we averaged over the full $\sim$12 Hr of observation and 2 GHz bandwidth, to ensure the best signal-to-noise ratio (SNR). We fit these images using the Common Astronomy Software Application (\textsc{casa}) \citep{McMullin:2007} tool \textsc{imstat}, which reports the statistics for a supplied region of a image. To determine if a source has detectable radio emission, we compared the measured peak flux density at the location of the source to the image RMS determined from a fit to a region centred on the source position with dimensions were 6 times the size of the restoring beam. We considered peak flux densities greater than 3 times the image RMS as detections. These fits were carried out in both the Stokes I and V images and the results are listed in Table \ref{table:emission-prop}. 

For sources observed with detectable radio emission in either of these two 2 GHz images, additional I and V images using 512~MHz frequency averaging were made to constrain the spectral index and polarisation frequency dependence. These 512~MHz images were then fit using the same method detailed above. By vector averaging the real components of the visibilities in time bins of 10 s, 30 s and 60 s over the two 2.0 GHz frequency bins, light curves were created to search for variability (section \ref{sec:var}).

\begin{figure*}
\includegraphics[width=5.5in]{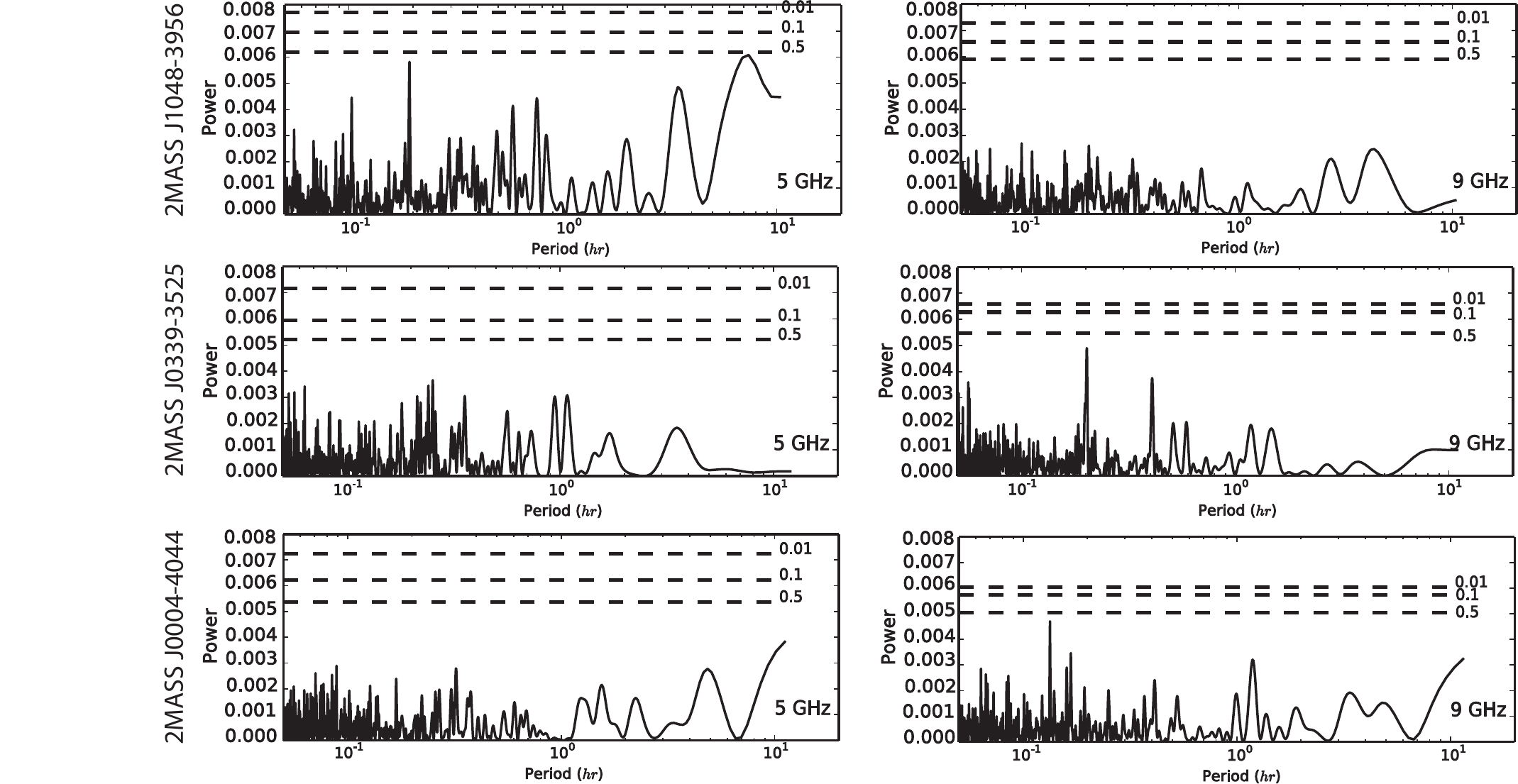}
\caption{Lomb-Scargle periodogram of the 10 s time averaged and 2~GHz frequency averaged, Stokes I flux values for  2\uppercase{mass}  J1048-3956 (top), 2\uppercase{mass}  J0339-3525 (middle), and 2\uppercase{mass}  J0004-4044 (bottom). The periodograms are calculated for both the 5 GHz (left column) and 9 GHz (right column) frequency bands. Dashed lines indicate false alarm probabilities of 0.01 (99$\%$), 0.1 (90$\%$), and 0.5 (61$\%$). We did not detect significant variability in these three sources.}
\label{fig:j0004-LS}
\end{figure*}

\section{Detections}\label{sec:results}

\begin{figure*}
\centering
\includegraphics[width=7.0in]{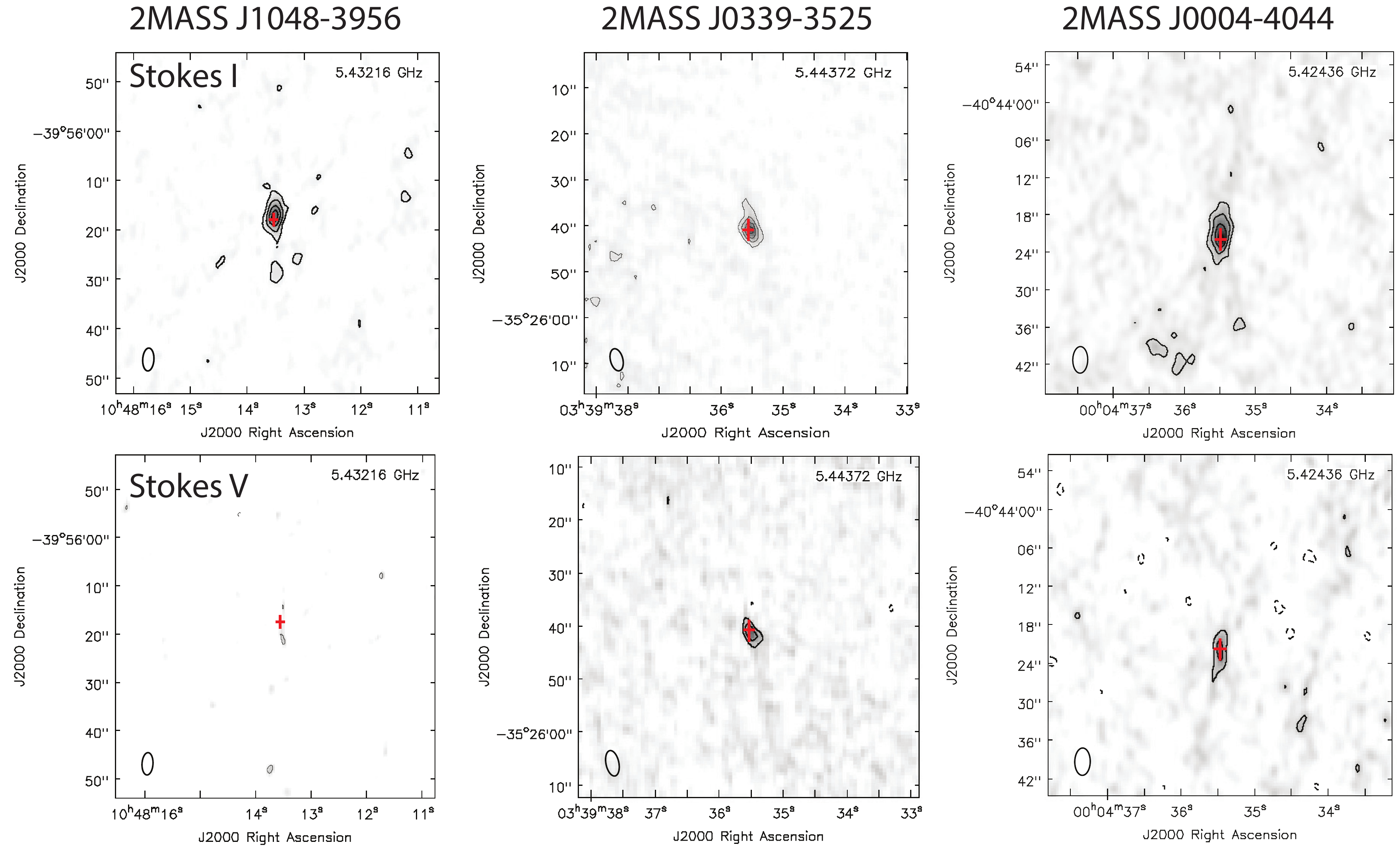}
\caption{Radio images for 2\uppercase{mass}  J1048-3956 (left-column), 2\uppercase{mass}  J0339-3525 (middle-column), and 2\uppercase{mass}  J0004-4044 (right-column) in Stokes I (top-row) and Stokes V (bottom-row) at 5.5~GHz. Additionally, contours are overlaid with levels at 3, 9, 18, 27  times the Stokes I RMS value of 8.2$\mu$Jy for 2\uppercase{mass}  J1048-3956 and 3, 6, 9, 12 20 times the Stokes I RMS values of 10.5$\mu$Jy and 8.3$\mu$Jy for 2\uppercase{mass}  J0339-3525 and 2\uppercase{mass}  J0004-4044, respectively.  For all sources the Stokes  V contours are  -10, -5, -3, 3, 5, 10 times the RMS values of 9.7$\mu$Jy for 2\uppercase{mass}  J1048-3956, 17.6$\mu$Jy for 2\uppercase{mass}  J0339-3525, and 13.3$\mu$Jy for 2\uppercase{mass}  J0004-4044. }
\label{fig:5GHz-Maps}
\end{figure*}

\begin{figure*}
\includegraphics[width=7.0in]{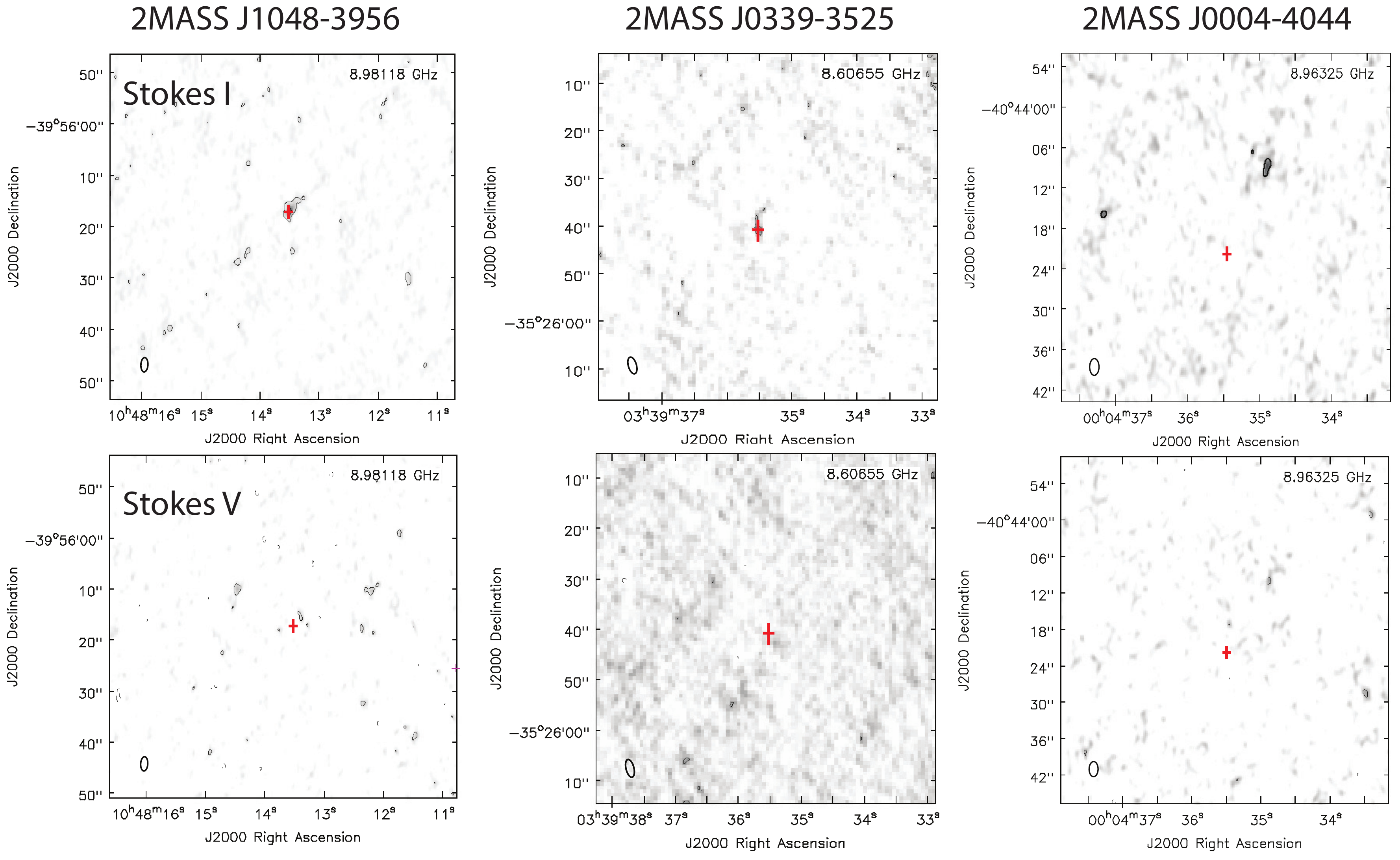}
\caption{Radio image for 2\uppercase{mass}  J1048-3956 (left-column), 2\uppercase{mass}  J0339-3525 (middle-column), and 2\uppercase{mass}  J0004-4044 (right-column) in Stokes I (top-row) and Stokes V (bottom-row) at 9.0~GHz. Additionally, contours are overlaid with levels at 3, 9, 18, 27  times the Stokes I RMS value of 10.6$\mu$Jy for 2\uppercase{mass}  J1048-3956 and 3, 6, 9, 12 20 times the Stokes I RMS values of 10.0$\mu$Jy and 8.2$\mu$Jy for 2\uppercase{mass}  J0339-3525 and 2\uppercase{mass}  J0004-4044, respectively.  For all sources the Stokes  V contours are  -10, -5, -3, 3, 5, 10 times the RMS values of 10.6$\mu$Jy for 2\uppercase{mass}  J1048-3956, 19.1$\mu$Jy for 2\uppercase{mass}  J0339-3525, and 12.8$\mu$Jy for 2\uppercase{mass}  J0004-4044. }
\label{fig:9GHz-Maps}
\end{figure*}

From fits to the 2~GHz frequency-averaged images, we find only three of our 15 target ultracool dwarfs have detectable levels of Stokes I emission in at least one of the two observing frequency bands, 2\uppercase{mass} J10481463--3956062 (hereafter 2\uppercase{mass}  J1048-3956), 2\uppercase{mass}I J0339352--352544 (hereafter 2\uppercase{mass}  J0339-3525), and 2\uppercase{mass}W J0004348--404405 (hereafter 2\uppercase{mass}  J0004-4044). The Stokes I and V images for these three sources at 5.5~GHz and 9.0~GHz are shown in Figures \ref{fig:5GHz-Maps}  and  \ref{fig:9GHz-Maps}, respectively. The flux density peaks in each of these images lie within the mean beam size of the predicted positions determined from the 2\uppercase{mass} astrometry \citep{Cutri:2003} and proper motion measurements from the literature \citep{Deacon:2005, Schmidt:2007, Faherty:2009}.

Both 2\uppercase{mass}  J1048-3956 and 2\uppercase{mass}  J0339-3525 have previous radio detections, while the detection of 2\uppercase{mass}  J0004-4044 is the first. Additionally, this survey includes radio limits on 4 sources, 2\uppercase{mass} I J0340094--672405, 2\uppercase{mass}  J02550357--4700509, 2\uppercase{mass} J03341218--4953322, and 2\uppercase{mass} J00244419--2708242,  with no previous radio observations. Combining our results with that of \citet{Antonova:2013}, \citet{Route:2013}, \citet{Burgasser:2013, Burgasser:2015}, and \citet{Kao:2015}, the number of ultracool dwarfs studied at radio radio frequencies is now 222, with 22 sources observed to have radio emission. From these numbers, $\sim$10$\%$ of ultracool dwarfs are observed to have radio emission. This is higher than the estimate by \citet{Antonova:2013} who get $\sim$6$\%$ when they consider only their observations and that of \citet{Mclean:2012}.  Additionally using this set of observations, \citet{Antonova:2013} note that the majority of the ultracool dwarfs with observed radio emission have a spectral type between M7 -- L3.5. However, this result may be due to the small number of observations of objects with spectral types >L3.5 included in their sample. If we add to the \citet{Antonova:2013} sample the results of \citet{Burgasser:2013}, \citet{Route:2013}, \citet{Kao:2015}, and our own radio detections, the fraction of ultracool dwarfs with observable radio emission remains constant ($\sim$10\%) across the spectral type range M7 -- T8.

\subsection{Variability }\label{sec:var}
To search for burst emission and any potential periodicity in the detected sources, we constructed light curves of the real visibilities in Stokes I and V for 2\uppercase{mass}  J1048-3956, 2\uppercase{mass}  J0339-3525, and 2\uppercase{mass}  J0004-4044. These light curves were made for 1 min, 0.5 min, and 0.1 min time averaged bins and 512 MHz, 1 GHz, and 2 GHz frequency averaged bins. In all combinations of averaging we do not detect any burst emission or variability in the quiescent component. 

We carried out a Lomb-Scargle analysis \citep{Lomb:1976, Scargle:1982} using the astroML scientific python modules \citep{VanderPlas:2012} to test the significance of the non-variability. The periodograms of the 3 sources with observed radio emission are shown in Figure \ref{fig:j0004-LS}. The dashed lines indicate false alarm probabilities for the 99$\%$, 90$\%$, and 61$\%$ levels. For all three source there are no significant peaks in the Lomb-Scargle power spectrum for variability on timescales less than an hour up to 12 hours.

\section{Characterising the Quiescent Emission}\label{sec:prop-emission}

2\uppercase{mass}  J1048-3956, 2\uppercase{mass}  J0339-3525 and 2\uppercase{mass}  J0004-4044 all have detectable levels of Stokes I emission at 5.5~GHz. For both 2\uppercase{mass}  J0339-3525 and 2\uppercase{mass}  J0004-4044 the 5.5~GHz emission is also observed to be polarised, with polarisation fractions of 0.53 for 2\uppercase{mass}  J0339-3525 and 0.44 for 2\uppercase{mass}  J0004-4044. At 9~GHz the radio emission from 2\uppercase{mass}  J0004-4044 is undetectable, however, for both 2\uppercase{mass}  J0339-3525 and 2\uppercase{mass}  J1048-3956 we still detect Stokes I emission from these sources. We do not detect any Stokes V emission for any of the sources at this higher frequency. 

\begin{table*}
	\centering
	\caption{Characteristics of the Radio Emission} \label{table:emission-prop}
	\begin{tabular}{llllllr}
		\hline
		2\uppercase{mass} Number & $S_{5.5 \text{GHz}}$($I$)\textsuperscript{a} & $S_{5.5\text{GHz}}$($V$)\textsuperscript{a} & $S_{9.0 \text{GHz}}$($I$)\textsuperscript{a} & $S_{9.0 \text{GHz}}$($V$)\textsuperscript{a}  & $\alpha_{4.7-9.7\ \text{GHz}}$ & T$_B$\\
 		& ($\mu$Jy) & ($\mu$Jy) & ($\mu$Jy) & ($\mu$Jy) & & (K)\\
		\hline
		10481258--1120082 & $<$44.7 & $<$26.7 & $<$34.5 & $<$31.8 & -- & $<$5.98$\times10^7$ \\
		14563831--2809473 & $<$29.9 & $<$29.0 & $<$37.4 & $<$36.8 & -- & $<$9.69$\times10^7$ \\
		11554286--2224586 & $<$31.2 & $<$25.8 & $<$34.8 & $<$34.2 & -- & $<$1.94$\times10^8$\\
		10481463--3956062 &  211.9$\pm$8.2 & $<$29.1 & 131.4$\pm$10.8 & $<$31.8 & -1.1$\pm$ 0.11 & 2.23$\times$10$^{8}$ \\ 
		00244419--2708242 & $<$111.0 & $<$37.8 & $<$237.0 &$<$48.9 & -- & $<$4.36$\times10^8$ \\
		0339352--352544 & 137.6$\pm$10.5 & 73.0$\pm$10.0 & 90.7$\pm$17.6 & $<$57.3 & -0.97$\pm$ 0.32 & 2.26$\times$10$^{8}$  \\ 
		03341218--4953322 & $<$29.4 & $<$26.1 & $<$37.8 & $<$42.0 & -- & $<$1.31$\times10^8$\\
		0853362--032932 & $<$42.9  & $<$35.4 & $<$53.7 & $<$50.4 & -- & $<$2.30$\times10^8$\\
		1507476--162738 & $<$37.6 & $<$28.5 & $<$36.6 &$<$35.4& -- & $<$3.58$\times10^8$  \\
		08354256--0819237 &  $<$32.0 &  $<$28.0 &  $<$42.0 &  $<$41.0 & -- &  $<$1.71$\times10^8$\\
		0004348--404405 & 100.0$\pm$8.3 & 44.3$\pm$8.2 & $<$39.9 & $<$38.4  & $<$-0.74 & 6.61$\times$10$^{8}$\\
		17502484--0016151 & $<$81 & $<$36 & $<$51 &$<$39&--& $<$3.43$\times10^8$\\
		0340094--672405 & $<$27.0 & $<$27.3 & $<$39.0 & $<$39.0 & -- & $<$1.75$\times10^8$\\
		02550357--4700509 & $<$30.9 & $<$26.7 & $<$34.6 & $<$31.8 & -- & $<$5.05$\times10^7$\\
		02572581--3105523 & $<$66.0 & $<$35.4 & $<$63.0 & $<$60.0 & -- & $<$4.02$\times10^8$\\
		\hline
		\multicolumn{7}{l}{\textsuperscript{a}\footnotesize{Upper limits listed are 3$\sigma$ limits based on the RMS values measured in the images for each source.}}
	\end{tabular}
\end{table*}

\subsection{Spectral Indices}
We did a least squares fit to the measured Stokes I values from the 512~MHz images to constrain spectral indices for the three detected sources (see Table \ref{table:emission-prop}) between 4.7 and 9.7~GHz. For all three sources the emission drops steeply with increasing frequency and shows no signs of a turn-over in the lower observing frequency band. 

Comparing the spectral indices we calculate for 2\uppercase{mass}  J0339-3525 and 2\uppercase{mass}  J1048-3956  to previously cited values, we find that they do not agree. For 2\uppercase{mass}  J0339-3525, \citet{Berger:2001} constrain the spectral index to be 2.1$\pm$0.3 between 4 and 8 GHz, indicating optically thick emission. The discrepancy between our result and that of \citet{Berger:2001} is most likely because \citet{Berger:2001} include both flare and quiescent emission in their analysis while we only consider the quiescent component. Yet note that the measured flux densities from \citet{Berger:2001} vary by more than a factor of 2 over the 3 months covered by their observations. This flux density variation could also be related to a variation in the spectral index.  The quiescent radio emission from 2\uppercase{mass}  J1048-3956 has been studied over a wide range of radio frequencies ($\sim$4.0-20~GHz) by \citet{Ravi:2011} who fit a spectral index of $\alpha$=1.71$\pm$0.09 to the observed radio emission. The smaller frequency coverage of our observations could be the cause of this difference. However, similar to 2\uppercase{mass}  J0339-3525,  the 4-9~GHz emission of 2\uppercase{mass}  J1048-3956 is variable on long-time scales where the radio emission previously observed by \citet{Ravi:2011} has a slightly higher flux density and is circularly polarised with a polarisation fraction of 0.25-0.4. Long-term variability in the measured flux densities and polarisation of UCDs has been observed in several other cases \citep{Mclean:2012} and may indicate a significant change in the physical characteristics of the emitting regions in these sources. Such variability is also well known in the case of radio flares from close stellar binaries and are attributed to changes in energisation \citep[e.g., RS CVns; ][]{Mutel:1998, Richards:2003}.

\subsection{Brightness Temperatures}
In order to assess the origin of the radio emission, we can calculate the brightness temperature of the observed emission. For a radio source at a  distance $d$ with an emitting volume of radius $R$, and flux density $S_{\nu}$ at frequency $\nu$ , the brightness temperature is given by,
\begin{equation}
T_b = 2.5\times10^{9}\left(\frac{S_{\nu}}{\text{mJy}}\right)\left(\frac{\nu}{\text{GHz}}\right)^{-2}\left(\frac{d}{\text{pc}}\right)^{2}\left(\frac{R}{R_{\text{J}}}\right)^{-2}\ \text{K},
\end{equation}
where $R_{J}$ is the radius of Jupiter and is the typical radius of very low mass stars and brown dwarfs \citep{Burrows:2001}. Assuming M-type stellar coronal dimensions of (1-2)$R_*$ \citep{Leto:2000}, the measured flux densities and appropriate upper limits for the non-detections imply brightness temperatures in the range of (0.5 - 6)$\times10^8$ K  at 5.5~GHz (see Table \ref{table:emission-prop}). 

\subsection{Origins for the emission}

\begin{figure*}
\includegraphics[width=7.0in]{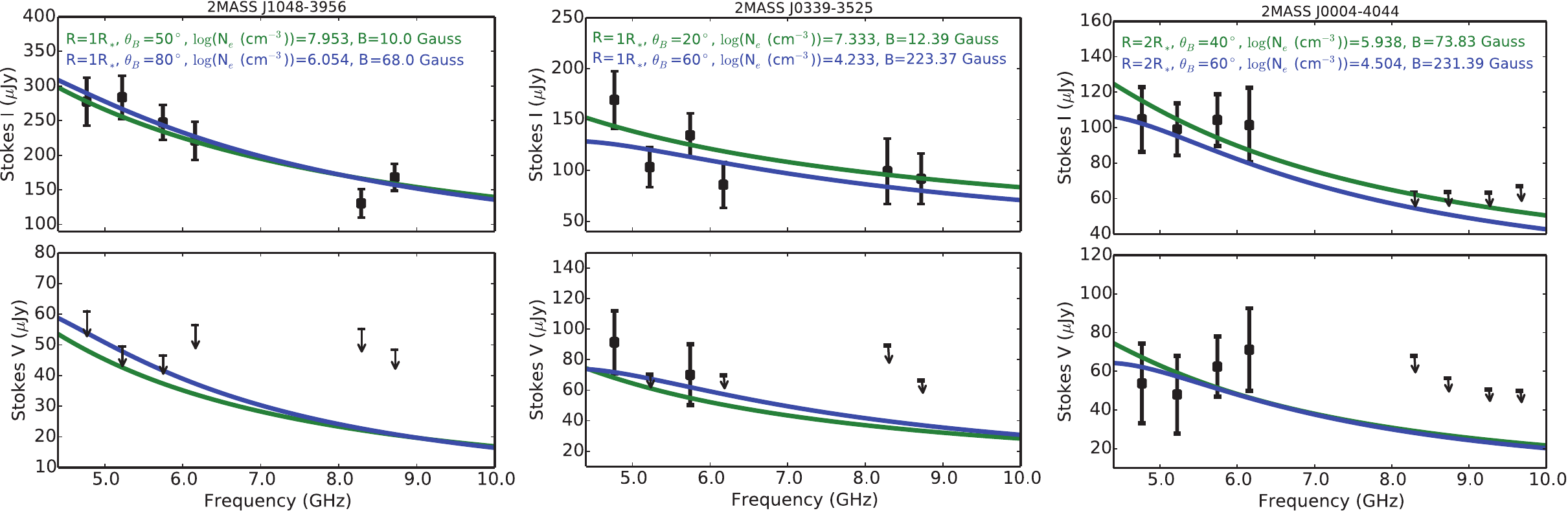}
\caption{Comparisons between two representative model curves and the measured Stokes I (top) and Stokes V (bottom) flux densities for 2\uppercase{mass} J1048-3956 (left), 2\uppercase{mass} J0339-3525 (middle), and 2\uppercase{mass} J0004-4044 (right). The measured values result from fits to the 512~MHz images for each source. The two different colours represent two different sets of model parameters that fall within the values listed in table \ref{table:model}. }
\label{fig:SEDModel}
\end{figure*}

The high brightness temperatures, combined with the spectral indices, and the measured circular polarisation of 2\uppercase{mass}  J0339-3525 and 2\uppercase{mass}  J0004-4044, rule out thermal bremsstrahlung emission and we find the emission to be more consistent with gyrosynchrotron emission from a non-thermal population of accelerated electrons \citep{Dulk:1985}. To model the observed emission characteristics, we used the expressions found in \citet{Robinson:1984} for the absorption and emission coefficients of gyrosynchrotron emission from mildly relativistic electrons with a power-law electron distribution given by, 
$$
N_e(E) = K E^{-\delta}$$ 
where $K =N_0(\delta-1){E_0}^{\delta-1}$ and  is $\delta$ the energy index, and with a low-energy cutoff $E_0$ = 10~keV. 

Such a model requires some knowledge of the magnetic geometry for the ultracool dwarf, specifically the angle of inclination between the line of sight and the magnetic axis, $\theta_B$. An estimate of this inclination angle can be obtained from observed variability in optical emission. However, 2\uppercase{mass}  J1048-3956, 2\uppercase{mass} J0339--3526, and 2\uppercase{mass}  J0004-4044 are observed to have little variability at these wavelengths \citep{Guenther:2009, Schmidt:2007, Stelzer:2012, Crossfield:2014}. Thus to characterize the plasma conditions responsible for this emission, we construct a simple coronal model consisting of a homogenous population of mildly-relativistic power-law electrons spiralling in a uniform magnetic field.  This simple model allows us to make an order-of-magnitude estimate for the power-law electron density and magnetic field strength without making assumptions concerning the geometry of the magnetic field. 

\begin{table}
	\centering
	\caption{Model Parameters} 
	\label{table:model}
	\begin{tabular}{lllll}
	\hline
		2\uppercase{mass} Number & $R$    & $\theta_B$ & $B$          & $\log$($N_e$) \\
 			   & ($R_*$)&                   &   (G) & (cm$^{-3}$)\\
		\hline
		10481463--3956062 & 1.0 -- 2.0 & 40$^{\circ}$ -- 80$^{\circ}$ & 10 -- 70  & 6.0 -- 8.0  \\
		0339352--352544 & 1.0 -- 2.0 & 20$^{\circ}$ -- 60$^{\circ}$ & 20 -- 223  & 4.1 -- 6.8  \\
		0004348--404405 & 1.5 -- 2.0 & 40$^{\circ}$ -- 60$^{\circ}$ & 73 -- 231  & 4.5 -- 6.2  \\
		\hline
	\end{tabular}
\end{table}

For this model we assumed the radius of the emitting region to range from (1-2)R$_*$, consistent with estimates for the emitting region dimension on M dwarfs \citep{Leto:2000}, and varied the emitting volume, the strength and orientation of the magnetic field, and the non-thermal electron density to best-fit the measured spectral energy distribution and fractional circular polarisation for each source. Since the spectral energy distribution for 2\uppercase{mass}  J1048-3956, 2\uppercase{mass}  J0339-3525, and 2\uppercase{mass}  J0004-4044 is consistent with optically thin gyrosynchrotron emission, we constrain the energy index $\delta$ for the model using the approximation \citep{Dulk:1985}
$$
\delta = (1.22-\alpha_{4.7 - 9.7  GHz} )/0.9 
$$
and the spectral indices given in Table \ref{table:emission-prop}. Additionally, the optically thin assumption implies that the gyrosynchrotron turn over frequency is less than 4.7~GHz and further limits the range of suitable model parameters.   

The model parameters that reproduce the observed Stokes I and V emission for 2\uppercase{mass}  J1048-3956, 2\uppercase{mass}  J0339-3525, and 2\uppercase{mass}  J0004-4044 are listed in Table \ref{table:model}.  Generally the range of values for the densities, magnetic field orientations and strengths, and the emission volumes are consistent with previous constraints found for ultracool dwarf magnetospheres \citep{Burgasser:2005, Osten:2006, Ravi:2011, Lynch:2015}. Sample SED's are shown in Figure~\ref{fig:SEDModel}, along with the corresponding observed fluxes in the 512~MHz images. The model parameters for each curve are annotated in the figure, where the colour of the text corresponds to the colour of the appropriate model curve. The SED fits are quite acceptable, with most of the model fluxes agreeing with upper limits and measured flux densities for both the Stokes I and V emission.

\section{Radio Emission Trends}\label{sec:general-trends}

%Previous radio surveys of ultracool dwarfs have found that the ratio of the radio to bolometric luminosity, $L_{rad}$/$L_{bol}$, increases with later spectral type. 
%Previous surveys to address the radio activity-rotation relation for M and L dwarfs found that 

\begin{figure}
\includegraphics[width=\columnwidth]{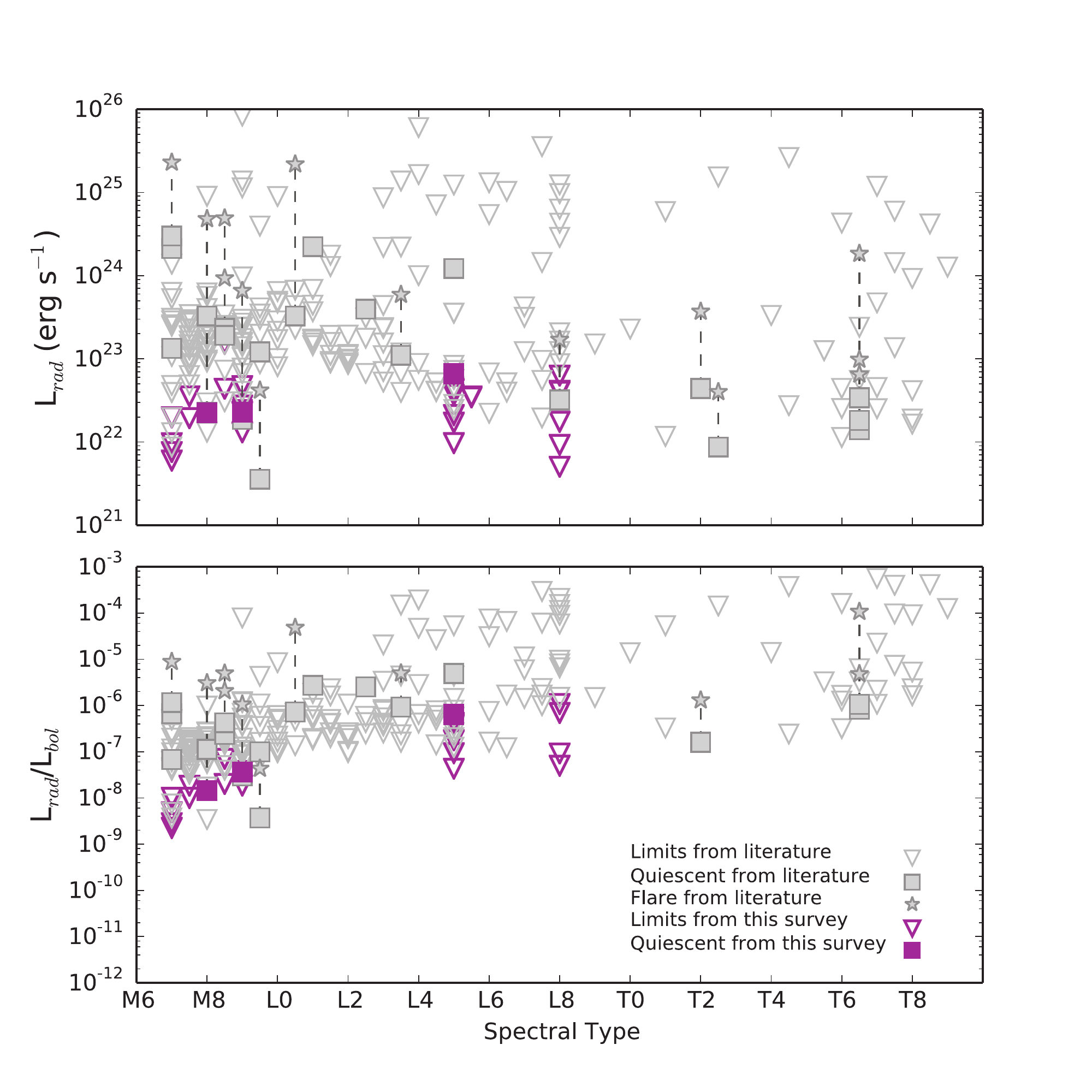}
\caption{(Top) Radio luminosity and (Bottom) ratio of the radio to bolometric luminosity as a function of spectral type for ultracool dwarfs. Shown are flares (stars), quiescent emission (squares) and upper limits (open triangles). The results from this survey are magenta and values from the literature \citep{Route:2013, Antonova:2013, Burgasser:2013, Burgasser:2015, Kao:2015} are grey. For objects with observed quiescent and flare emission, the values are connected by dashed lines. }
\label{fig:LumSpec}
\end{figure}

\begin{figure}
\includegraphics[width=\columnwidth]{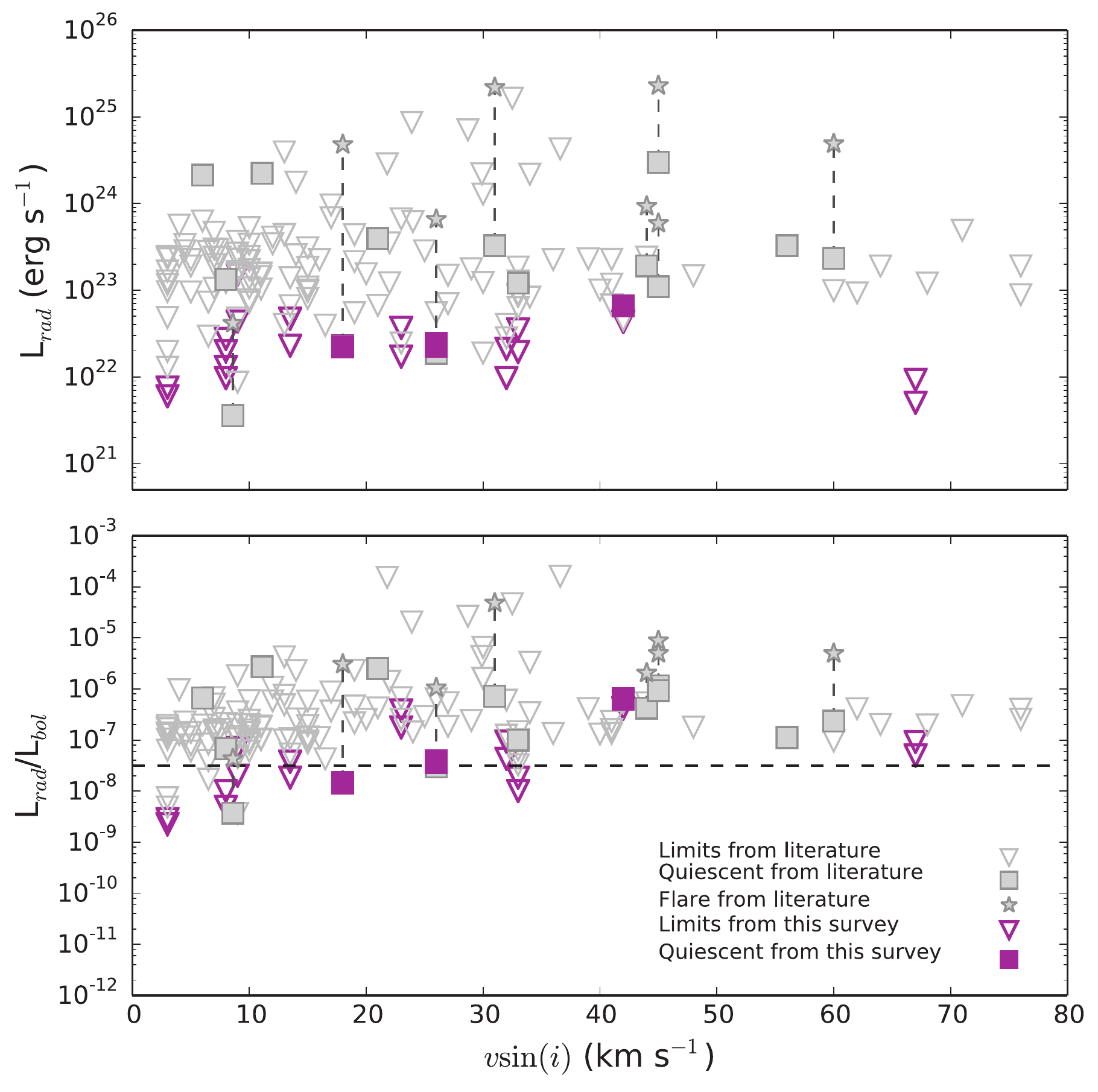}
\caption{(Top) Radio luminosity and (Bottom) ratio of the radio to bolometric luminosity as a function of projected rotational velocity for ultracool dwarfs. Upper limits (open triangles), flares (stars) and quiescent emission (squares) are shown for both this survey (magenta) and the literature \citep[grey;][]{Route:2013, Antonova:2013, Burgasser:2013, Burgasser:2015}.  The dashed line in the bottom panel indicates the radio activity-rotation saturation level for early type M dwarfs from \citet{Mclean:2012}.}
\label{fig:vsin}
\end{figure}

To put the results of our survey into context, we compare our radio results with that of previous radio surveys in Figures \ref{fig:LumSpec} and \ref{fig:vsin}. In these figures our measurements are the magenta points and the grey points are from \citet{Route:2013}, \citet{Antonova:2013}, \citet{Burgasser:2013, Burgasser:2015}, and \citet{Kao:2015}, where appropriate.  In these figures we include upper limits (open triangles), quiescent emission (squares) and flares (stars). We find that our upper limits on the radio luminosity are lower than those placed by the previous surveys but are still comparable to the detected quiescent emission from the least bright sources. The observed luminosities for 2\uppercase{mass}  J1048-3956 and 2\uppercase{mass}  J0339-3525 are consistent with those from previous surveys.

As a function of spectral type (Figure \ref{fig:LumSpec}), we observe the radio luminosity to be constant, agreeing with previous radio surveys which found $L_{rad}\sim10^{23\pm0.5}$ erg s$^{-1}$ for objects with spectral type M0-L5 \citep{Berger:2010}.  Furthermore, when we compare the ratio of radio to bolometric luminosity to spectral type, we observe the previously noted trend of increased activity with later spectral type. This is in contrast with observations of other activity tracers, such as $H\alpha$ and X-ray emission, where this ratio decreases past spectral type M7 \citep{Berger:2010}. 

Figure \ref{fig:vsin} shows the radio luminosity and ratio of radio to bolometric luminosity as a function of rotation rate. 2\uppercase{mass} J1048$-$3956, 2\uppercase{mass}  J0339-3525, and 2\uppercase{mass}  J0004-4044 are all rapid rotators with $v\sin(i)\gtrsim20$ km s$^{-1}$. The radio luminosities for these three sources are observed to be fairly constant with rotation rate agreeing with the previous results of \citet{Mclean:2012}. However, note that rapid rotation does not necessarily indicate a source will have observable radio emission. In fact the most rapidly rotating source in our sample of 15, 2\uppercase{mass} J02550357--4700509, has an upper limit of $L_{rad}\sim9\times10^{11}$~erg s$^{-1}$, which is lower than the luminosity of the detected sources. 

If we look at the ratio of the radio to bolometric luminosity for this sample of sources we see that 2\uppercase{mass}  J1048-3956 and 2\uppercase{mass}  J0339-3525 have ratios that lie along the radio activity-rotation saturation level observed in early-M dwarfs, $L_{rad}/L_{bol}\sim10^{-7.5}$ \citep{Mclean:2012}. However, note that sources with rotation rates $\gtrsim$30 km s$^{-1}$, including our observation of 2\uppercase{mass}  J0004-4044, all lie above the early-M dwarf saturation level. As noted by \citet{Mclean:2012} these more rapidly rotating sources appear to tend toward the higher ratio of radio to bolometric luminosity of  $L_{rad}/L_{bol}\sim10^{-6.5}$.

\subsection{X-ray/Radio Correlation}
As mentioned in section \ref{sec:intro}, there is a tight correlation between the radio and X-ray emission for coronally active stars ranging from spectral type F to mid-M, where $L_{\nu}$/$L_x$$\sim10^{-15.5}$Hz$^{-1}$ \citep{Gudel:1993a, Benz:1994}. The first radio observation of an ultracool dwarf found $L_{\nu}$/$L_x$$\sim10^{-11.5}$Hz$^{-1}$ for this source \citep{Berger:2001}, suggesting that the GB relation may be severely violated by these objects. Subsequent radio detections of ultracool dwarfs have found that these objects display a wide range of behaviour with regard to the GB relation, where some sources are strongly radio overluminous varying from the GB relation by several orders of magnitude, while others could be consistent with the this relation \citep{Williams:2014}.

\begin{figure}
\includegraphics[width=3.0in]{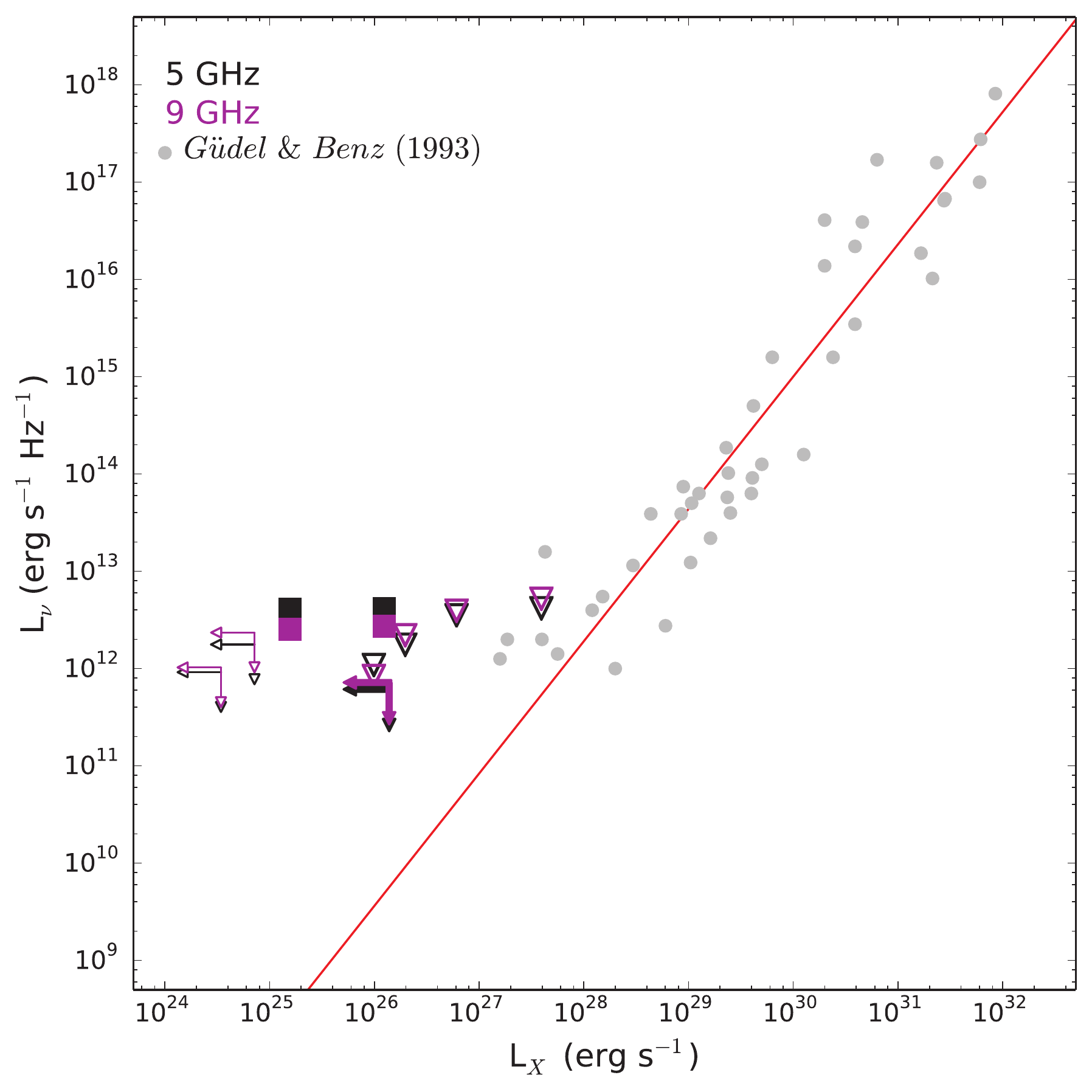}
\caption{Quiescent radio luminosity as a function of X-ray luminosity for a wide range of coronally active stars. The grey points are the results from \citet{Gudel:1993a} and the red-line is a fit to these data from \citet{Berger:2010}. The results from this survey are given by the black (5 GHz) and magenta (9 GHz) data points. The solid arrows represent the upper limits from stacking the observations with non-detections. }
\label{fig:GBRel}
\end{figure}

Using the radio luminosities, $L_{\nu}$, for our 15 sources as well as X-ray luminosities, $L_x$, from the literature, we can determine if these sources fall along the GB relation. Figure \ref{fig:GBRel} shows this comparison, where we have also plotted the observed data from \citet{Gudel:1993a} (grey points) and the linear fit of 
$\log(L_{\nu}) = 1.36\left[\log(L_x) - 18.9\right]$
to this data \citep{Berger:2010}. The scatter of the data from \citet{Gudel:1993a} around this line is 0.6 dex when we measure the deviation at a fixed $L_x$. The relative scatter of the data to the best fit line (i.e. measured perpendicular to the line) is 0.2 dex. In order to be consistent with the analysis of \citet{Williams:2014}, we define the difference between the measured ratio of radio to X-ray luminosity and the GB relation as the perpendicular distance between the measured value and the best fit line. 

Out of the three sources in our survey with detectable levels of radio emission only 2\uppercase{mass}  J1048-3956 and 2\uppercase{mass}  J0339-3525 have measured X-ray luminosities in the literature. Comparing the ratio of the radio to X-ray luminosity for these two sources we find that they differ from the GB relation by 2.5 dex and 1.7 dex, for 2\uppercase{mass}  J1048-3956 and 2\uppercase{mass}  J0339-3525 respectively.  The variation from the GB relation for these two sources, while significant, is not nearly as extreme as the variation observed for other ultracool dwarfs \citep[e.g. TVLM-513; ][]{Berger:2008a, Williams:2014}. 

Most of our measured radio upper limits are within 1.5 dex of the GB relation. Given these upper limits the actual radio luminosity of these sources has the potential to be consistent with the GB relation. However for 2 of our objects, 2\uppercase{mass} J1507476--162738 and 2\uppercase{mass} J02550357--4700509, their measured radio upper limits place them $\sim$2.6 dex away from the GB relation. This indicates that the measured radio luminosities for these two sources would have to be significantly less than the upper limits in order for them to be consistent with the GB relation.

Following the method outlined in \citet{Hancock:2011}, we stacked 11 of the 12 observations with non-detections. We excluded the observation of 2\uppercase{mass} J00244419--2708242 because we were unable to fully remove a bright field source located at R.A. = 00:24:36.077 Decl.=-27.07.44.64. In the stacked image we do not detect any radio emission and measure upper limits of 12$\mu$Jy (5.5 GHz) and 14.1$\mu$Jy (9.0 GHz). Noting that the X-ray luminosities for these sources are $L_{x}/L_{bol}$>-5 and calculating the effective distance for the staked image to be 6.5 pc, we plot this upper limit in Figure \ref{fig:GBRel} (solid arrows). Similar to the other upper limits from this survey, this upper limit is ~1.3 dex from the GB relation. 

\section{Summary}

From a sample of 15 late-M and L dwarfs, located within 10 pc of the Sun, we have detected quiescent radio emission at frequencies between 4.7 and 9.7~GHz for three sources, 2\uppercase{mass}  J1048-3956, 2\uppercase{mass}  J0339-3525, and 2\uppercase{mass}  J0004-4044. While both 2\uppercase{mass}  J1048-3956 and 2\uppercase{mass}  J0339-3525 have previous radio detections in the literature, this is the first detection of radio emission from 2\uppercase{mass}  J0004-4044. Additionally, we place the first upper limits on the radio emission from 2\uppercase{mass} J0024-2708, 2\uppercase{mass} J03341218--4953322, 2\uppercase{mass} J0340-6724, and 2\uppercase{mass} J02550357--4700509. This increases the number of ultracool dwarfs studied at radio frequencies to 216, with 17 sources with observed radio emission.

We find that the observed Stokes I and V radio emission from 2\uppercase{mass}  J1048-3956, 2\uppercase{mass}  J0339-3525, and 2\uppercase{mass}W J0004348-404405 is well modelled by optically thin gyrosynchrotron emission from a homogenous population of power-law electrons, with density between 10$^4$ -- 10$^8$ cm$^{-3}$, spiralling in a magnetic field with a strength of 10 -- 300 Gauss and orientation between 40$^{\circ}$--80$^{\circ}$. These parameter ranges are still very large and could benefit from an observation of the spectral turn-over frequency.  In the case of gyrosynchrotron emission this frequency is dependent on the electron density and magnetic field orientation and strength \citep{Dulk:1985}. Here we assume that this frequency is $<$4~GHz based on the observed spectral energy distributions, however an actual measurement will place stronger constraints on the model parameters.

We also compare the general emission trends of our sample of 15 sources to the results of previous surveys of ultracool dwarfs. As observed in previous radio studies of ultracool dwarfs, we find the observed radio luminosities to be constant with both spectral type and rotation rate. We also find that the ratio of radio to bolometric luminosity to increase towards later type objects and higher rotational velocities.
Additionally, using X-ray luminosities from the literature we find that the ratio of radio to X-ray luminosity for 2\uppercase{mass}  J1048-3956 and 2\uppercase{mass}  J0339-3525 vary significantly from the GB relation. The majority of the radio upper limits are $\lesssim$1.5 dex from the GB relation, however for two source the upper limits differ from GB relation by more than 2 dex. For these two sources the measured radio luminosity would have to be much lower than these upper limits to be consistent with the GB relation.

\section*{Acknowledgements}

%The Acknowledgements section is not numbered. Here you can thank helpful
%colleagues, acknowledge funding agencies, telescopes and facilities used etc.
The Australia Telescope Compact Array is part of the Australia Telescope which is funded by the Commonwealth of Australia for operation as a National Facility managed by CSIRO. Parts of this research were conducted by the Australian Research Council Centre of Excellence for All-sky Astrophysics (CAASTRO), through project number CE110001020.

%%%%%%%%%%%%%%%%%%%%%%%%%%%%%%%%%%%%%%%%%%%%%%%%%%



\bibliographystyle{mnras}
\bibliography{ucd-papers} % if your bibtex file is called example.bib

\begin{thebibliography}{}
\makeatletter
\relax
\def\mn@urlcharsother{\let\do\@makeother \do\$\do\&\do\#\do\^\do\_\do\%\do\~}
\def\mn@doi{\begingroup\mn@urlcharsother \@ifnextchar [ {\mn@doi@}
  {\mn@doi@[]}}
\def\mn@doi@[#1]#2{\def\@tempa{#1}\ifx\@tempa\@empty \href
  {http://dx.doi.org/#2} {doi:#2}\else \href {http://dx.doi.org/#2} {#1}\fi
  \endgroup}
\def\mn@eprint#1#2{\mn@eprint@#1:#2::\@nil}
\def\mn@eprint@arXiv#1{\href {http://arxiv.org/abs/#1} {{\tt arXiv:#1}}}
\def\mn@eprint@dblp#1{\href {http://dblp.uni-trier.de/rec/bibtex/#1.xml}
  {dblp:#1}}
\def\mn@eprint@#1:#2:#3:#4\@nil{\def\@tempa {#1}\def\@tempb {#2}\def\@tempc
  {#3}\ifx \@tempc \@empty \let \@tempc \@tempb \let \@tempb \@tempa \fi \ifx
  \@tempb \@empty \def\@tempb {arXiv}\fi \@ifundefined
  {mn@eprint@\@tempb}{\@tempb:\@tempc}{\expandafter \expandafter \csname
  mn@eprint@\@tempb\endcsname \expandafter{\@tempc}}}

\bibitem[\protect\citeauthoryear{{Allred}, {Hawley}, {Abbett}  \&
  {Carlsson}}{{Allred} et~al.}{2006}]{Allred:2006}
{Allred} J.~C.,  {Hawley} S.~L.,  {Abbett} W.~P.,   {Carlsson} M.,  2006,
  \mn@doi [\apj] {10.1086/503314}, \href
  {http://adsabs.harvard.edu/abs/2006ApJ...644..484A} {644, 484}

\bibitem[\protect\citeauthoryear{{Antonova}, {Hallinan}, {Doyle}, {Yu},
  {Kuznetsov}, {Metodieva}, {Golden}  \& {Cruz}}{{Antonova}
  et~al.}{2013}]{Antonova:2013}
{Antonova} A.,  {Hallinan} G.,  {Doyle} J.~G.,  {Yu} S.,  {Kuznetsov} A.,
  {Metodieva} Y.,  {Golden} A.,   {Cruz} K.~L.,  2013, \mn@doi [\aap]
  {10.1051/0004-6361/201118583}, \href
  {http://adsabs.harvard.edu/abs/2013A%26A...549A.131A} {549, A131}

\bibitem[\protect\citeauthoryear{{Benz} \& {Guedel}}{{Benz} \&
  {Guedel}}{1994}]{Benz:1994}
{Benz} A.~O.,  {Guedel} M.,  1994, \aap, \href
  {http://adsabs.harvard.edu/abs/1994A%26A...285..621B} {285, 621}

\bibitem[\protect\citeauthoryear{Berger}{Berger}{2002}]{Berger:2002}
Berger E.,  2002, \apj, 572, 503

\bibitem[\protect\citeauthoryear{Berger}{Berger}{2006}]{Berger:2006}
Berger E.,  2006, \apj, 648, 629

\bibitem[\protect\citeauthoryear{Berger et~al.,}{Berger
  et~al.}{2001}]{Berger:2001}
Berger E.,  et~al., 2001, Nature, 410, 338

\bibitem[\protect\citeauthoryear{Berger et~al.,}{Berger
  et~al.}{2008}]{Berger:2008a}
Berger E.,  et~al., 2008, \apj, 673, 1080

\bibitem[\protect\citeauthoryear{Berger et~al.,}{Berger
  et~al.}{2009}]{Berger:2009}
Berger E.,  et~al., 2009, \apj, 695, 310

\bibitem[\protect\citeauthoryear{{Berger} et~al.,}{{Berger}
  et~al.}{2010}]{Berger:2010}
{Berger} E.,  et~al., 2010, \mn@doi [\apj] {10.1088/0004-637X/709/1/332}, \href
  {http://adsabs.harvard.edu/abs/2010ApJ...709..332B} {709, 332}

\bibitem[\protect\citeauthoryear{{Burgasser} \& {Putman}}{{Burgasser} \&
  {Putman}}{2005}]{Burgasser:2005}
{Burgasser} A.~J.,  {Putman} M.~E.,  2005, \mn@doi [\apj] {10.1086/429788},
  \href {http://adsabs.harvard.edu/abs/2005ApJ...626..486B} {626, 486}

\bibitem[\protect\citeauthoryear{{Burgasser}, {Melis}, {Zauderer}  \&
  {Berger}}{{Burgasser} et~al.}{2013}]{Burgasser:2013}
{Burgasser} A.~J.,  {Melis} C.,  {Zauderer} B.~A.,   {Berger} E.,  2013,
  \mn@doi [\apjl] {10.1088/2041-8205/762/1/L3}, \href
  {http://adsabs.harvard.edu/abs/2013ApJ...762L...3B} {762, L3}

\bibitem[\protect\citeauthoryear{{Burgasser}, {Melis}, {Todd}, {Gelino},
  {Hallinan}  \& {Bardalez Gagliuffi}}{{Burgasser}
  et~al.}{2015}]{Burgasser:2015}
{Burgasser} A.~J.,  {Melis} C.,  {Todd} J.,  {Gelino} C.~R.,  {Hallinan} G.,
  {Bardalez Gagliuffi} D.,  2015, \mn@doi [\aj] {10.1088/0004-6256/150/6/180},
  \href {http://adsabs.harvard.edu/abs/2015AJ....150..180B} {150, 180}

\bibitem[\protect\citeauthoryear{{Burrows}, {Hubbard}, {Lunine}  \&
  {Liebert}}{{Burrows} et~al.}{2001}]{Burrows:2001}
{Burrows} A.,  {Hubbard} W.~B.,  {Lunine} J.~I.,   {Liebert} J.,  2001, \mn@doi
  [Reviews of Modern Physics] {10.1103/RevModPhys.73.719}, \href
  {http://adsabs.harvard.edu/abs/2001RvMP...73..719B} {73, 719}

\bibitem[\protect\citeauthoryear{{Crossfield}}{{Crossfield}}{2014}]{Crossfield:2014}
{Crossfield} I.~J.~M.,  2014, \mn@doi [\aap] {10.1051/0004-6361/201423750},
  \href {http://adsabs.harvard.edu/abs/2014A%26A...566A.130C} {566, A130}

\bibitem[\protect\citeauthoryear{{Cutri} et~al.,}{{Cutri}
  et~al.}{2003}]{Cutri:2003}
{Cutri} R.~M.,  et~al., 2003, VizieR Online Data Catalog, \href
  {http://adsabs.harvard.edu/abs/2003yCat.2246....0C} {2246, 0}

\bibitem[\protect\citeauthoryear{{Deacon}, {Hambly}  \& {Cooke}}{{Deacon}
  et~al.}{2005}]{Deacon:2005}
{Deacon} N.~R.,  {Hambly} N.~C.,   {Cooke} J.~A.,  2005, \mn@doi [\aap]
  {10.1051/0004-6361:20042002}, \href
  {http://adsabs.harvard.edu/abs/2005A%26A...435..363D} {435, 363}

\bibitem[\protect\citeauthoryear{Dulk}{Dulk}{1985}]{Dulk:1985}
Dulk G.~A.,  1985, \mn@doi [\araa] {10.1146/annurev.aa.23.090185.001125}, \href
  {http://adsabs.harvard.edu/abs/1985ARA%26A..23..169D} {23, 169}

\bibitem[\protect\citeauthoryear{{Faherty}, {Burgasser}, {Cruz}, {Shara},
  {Walter}  \& {Gelino}}{{Faherty} et~al.}{2009}]{Faherty:2009}
{Faherty} J.~K.,  {Burgasser} A.~J.,  {Cruz} K.~L.,  {Shara} M.~M.,  {Walter}
  F.~M.,   {Gelino} C.~R.,  2009, \mn@doi [\aj] {10.1088/0004-6256/137/1/1},
  \href {http://adsabs.harvard.edu/abs/2009AJ....137....1F} {137, 1}

\bibitem[\protect\citeauthoryear{{Gizis}, {Monet}, {Reid}, {Kirkpatrick},
  {Liebert}  \& {Williams}}{{Gizis} et~al.}{2000}]{Gizis:2000}
{Gizis} J.~E.,  {Monet} D.~G.,  {Reid} I.~N.,  {Kirkpatrick} J.~D.,  {Liebert}
  J.,   {Williams} R.~J.,  2000, \mn@doi [\aj] {10.1086/301456}, \href
  {http://adsabs.harvard.edu/abs/2000AJ....120.1085G} {120, 1085}

\bibitem[\protect\citeauthoryear{G\"udel \& Benz}{G\"udel \&
  Benz}{1993}]{Gudel:1993a}
G\"udel M.,  Benz A.~O.,  1993, \apj, 405, L63

\bibitem[\protect\citeauthoryear{{Guenther}, {Zapatero Osorio}, {Mehner}  \&
  {Mart{\'{\i}}n}}{{Guenther} et~al.}{2009}]{Guenther:2009}
{Guenther} E.~W.,  {Zapatero Osorio} M.~R.,  {Mehner} A.,   {Mart{\'{\i}}n}
  E.~L.,  2009, \mn@doi [\aap] {10.1051/0004-6361/200810216}, \href
  {http://adsabs.harvard.edu/abs/2009A%26A...498..281G} {498, 281}

\bibitem[\protect\citeauthoryear{Hallinan, Antonova, Doyle, Bourke, Brisken  \&
  Golden}{Hallinan et~al.}{2006}]{Hallinan:2006}
Hallinan G.,  Antonova A.,  Doyle J.~G.,  Bourke S.,  Brisken W.~F.,   Golden
  A.,  2006, \apj, 653, 690

\bibitem[\protect\citeauthoryear{Hallinan et~al.,}{Hallinan
  et~al.}{2007}]{Hallinan:2007}
Hallinan G.,  et~al., 2007, \apj, 663, L25

\bibitem[\protect\citeauthoryear{{Hallinan} et~al.,}{{Hallinan}
  et~al.}{2015}]{Hallinan:2015}
{Hallinan} G.,  et~al., 2015, \mn@doi [\nat] {10.1038/nature14619}, \href
  {http://adsabs.harvard.edu/abs/2015Natur.523..568H} {523, 568}

\bibitem[\protect\citeauthoryear{{Hancock}, {Gaensler}  \& {Murphy}}{{Hancock}
  et~al.}{2011}]{Hancock:2011}
{Hancock} P.~J.,  {Gaensler} B.~M.,   {Murphy} T.,  2011, \mn@doi [\apjl]
  {10.1088/2041-8205/735/2/L35}, \href
  {http://adsabs.harvard.edu/abs/2011ApJ...735L..35H} {735, L35}

\bibitem[\protect\citeauthoryear{{Hawley}, {Gizis}  \& {Reid}}{{Hawley}
  et~al.}{1996}]{Hawley:1996}
{Hawley} S.~L.,  {Gizis} J.~E.,   {Reid} I.~N.,  1996, \mn@doi [\aj]
  {10.1086/118222}, \href {http://adsabs.harvard.edu/abs/1996AJ....112.2799H}
  {112, 2799}

\bibitem[\protect\citeauthoryear{{Kao}, {Hallinan}, {Pineda}, {Escala},
  {Burgasser}, {Bourke}  \& {Stevenson}}{{Kao} et~al.}{2015}]{Kao:2015}
{Kao} M.~M.,  {Hallinan} G.,  {Pineda} J.~S.,  {Escala} I.,  {Burgasser} A.,
  {Bourke} S.,   {Stevenson} D.,  2015, preprint, \href
  {http://adsabs.harvard.edu/abs/2015arXiv151103661K} {} (\mn@eprint {arXiv}
  {1511.03661})

\bibitem[\protect\citeauthoryear{{Leto}, {Pagano}, {Linsky}, {Rodon{\`o}}  \&
  {Umana}}{{Leto} et~al.}{2000}]{Leto:2000}
{Leto} G.,  {Pagano} I.,  {Linsky} J.~L.,  {Rodon{\`o}} M.,   {Umana} G.,
  2000, \aap, \href {http://adsabs.harvard.edu/abs/2000A%26A...359.1035L} {359,
  1035}

\bibitem[\protect\citeauthoryear{{Lomb}}{{Lomb}}{1976}]{Lomb:1976}
{Lomb} N.~R.,  1976, \mn@doi [\apss] {10.1007/BF00648343}, \href
  {http://adsabs.harvard.edu/abs/1976Ap%26SS..39..447L} {39, 447}

\bibitem[\protect\citeauthoryear{{Lynch}, {Mutel}  \& {G{\"u}del}}{{Lynch}
  et~al.}{2015}]{Lynch:2015}
{Lynch} C.,  {Mutel} R.~L.,   {G{\"u}del} M.,  2015, \mn@doi [\apj]
  {10.1088/0004-637X/802/2/106}, \href
  {http://adsabs.harvard.edu/abs/2015ApJ...802..106L} {802, 106}

\bibitem[\protect\citeauthoryear{{Machado}, {Avrett}, {Vernazza}  \&
  {Noyes}}{{Machado} et~al.}{1980}]{Machado:1980}
{Machado} M.~E.,  {Avrett} E.~H.,  {Vernazza} J.~E.,   {Noyes} R.~W.,  1980,
  \mn@doi [\apj] {10.1086/158467}, \href
  {http://adsabs.harvard.edu/abs/1980ApJ...242..336M} {242, 336}

\bibitem[\protect\citeauthoryear{McLean, Berger, Irwin, Forbrich  \&
  Reiners}{McLean et~al.}{2011}]{Mclean:2011}
McLean M.,  Berger E.,  Irwin J.,  Forbrich J.,   Reiners A.,  2011, \apj, 741,
  27M

\bibitem[\protect\citeauthoryear{{McLean}, {Berger}  \& {Reiners}}{{McLean}
  et~al.}{2012}]{Mclean:2012}
{McLean} M.,  {Berger} E.,   {Reiners} A.,  2012, \mn@doi [\apj]
  {10.1088/0004-637X/746/1/23}, \href
  {http://adsabs.harvard.edu/abs/2012ApJ...746...23M} {746, 23}

\bibitem[\protect\citeauthoryear{{McMullin}, {Waters}, {Schiebel}, {Young}  \&
  {Golap}}{{McMullin} et~al.}{2007}]{McMullin:2007}
{McMullin} J.~P.,  {Waters} B.,  {Schiebel} D.,  {Young} W.,   {Golap} K.,
  2007, in {Shaw} R.~A.,  {Hill} F.,   {Bell} D.~J.,  eds,  Astronomical
  Society of the Pacific Conference Series Vol. 376, Astronomical Data Analysis
  Software and Systems XVI. p.~127

\bibitem[\protect\citeauthoryear{{Mohanty}, {Basri}, {Shu}, {Allard}  \&
  {Chabrier}}{{Mohanty} et~al.}{2002}]{Mohanty:2002}
{Mohanty} S.,  {Basri} G.,  {Shu} F.,  {Allard} F.,   {Chabrier} G.,  2002,
  \mn@doi [\apj] {10.1086/339911}, \href
  {http://adsabs.harvard.edu/abs/2002ApJ...571..469M} {571, 469}

\bibitem[\protect\citeauthoryear{Mutel, Molnar, Waltman  \& Ghigo}{Mutel
  et~al.}{1998}]{Mutel:1998}
Mutel R.~L.,  Molnar L.~A.,  Waltman E.~B.,   Ghigo F.~D.,  1998, \apj, 507,
  371

\bibitem[\protect\citeauthoryear{{Neuh{\"a}user} et~al.,}{{Neuh{\"a}user}
  et~al.}{1999}]{Neuhauser:1999}
{Neuh{\"a}user} R.,  et~al., 1999, \aap, \href
  {http://adsabs.harvard.edu/abs/1999A%26A...343..883N} {343, 883}

\bibitem[\protect\citeauthoryear{{Neupert}}{{Neupert}}{1968}]{Neupert:1968}
{Neupert} W.~M.,  1968, \mn@doi [\apjl] {10.1086/180220}, \href
  {http://adsabs.harvard.edu/abs/1968ApJ...153L..59N} {153, L59}

\bibitem[\protect\citeauthoryear{Osten \& Jayawardhana}{Osten \&
  Jayawardhana}{2006}]{Osten:2006a}
Osten R.~A.,  Jayawardhana R.,  2006, \apj, 644, L67

\bibitem[\protect\citeauthoryear{{Osten} \& {Wolk}}{{Osten} \&
  {Wolk}}{2009}]{Osten:2009}
{Osten} R.~A.,  {Wolk} S.~J.,  2009, \mn@doi [\apj]
  {10.1088/0004-637X/691/2/1128}, \href
  {http://adsabs.harvard.edu/abs/2009ApJ...691.1128O} {691, 1128}

\bibitem[\protect\citeauthoryear{Osten, Hawley, Bastian  \& Reid}{Osten
  et~al.}{2006a}]{Osten:2006b}
Osten R.~A.,  Hawley S.~L.,  Bastian T.~S.,   Reid I.~N.,  2006a, \apj, 637,
  518

\bibitem[\protect\citeauthoryear{Osten, Hawley, Allred, Johns-Krull, Brown  \&
  Harper}{Osten et~al.}{2006b}]{Osten:2006}
Osten R.~A.,  Hawley S.~L.,  Allred J.,  Johns-Krull C.~M.,  Brown A.,   Harper
  G.~M.,  2006b, \apj, 647, 1349

\bibitem[\protect\citeauthoryear{Phan-Bao, Osten, Lim, Mart\`{i}n  \&
  Ho}{Phan-Bao et~al.}{2007}]{Phan-Bao:2007}
Phan-Bao N.,  Osten R.~A.,  Lim J.,  Mart\`{i}n E.~L.,   Ho P. T.~P.,  2007,
  \apj, 658, 553

\bibitem[\protect\citeauthoryear{{Ravi}, {Hallinan}, {Hobbs}  \&
  {Champion}}{{Ravi} et~al.}{2011}]{Ravi:2011}
{Ravi} V.,  {Hallinan} G.,  {Hobbs} G.,   {Champion} D.~J.,  2011, \mn@doi
  [\apj] {10.1088/2041-8205/735/1/L2}, \href
  {http://adsabs.harvard.edu/abs/2011ApJ...735L...2R} {735, L2}

\bibitem[\protect\citeauthoryear{{Reid}, {Cruz}, {Kirkpatrick}, {Allen},
  {Mungall}, {Liebert}, {Lowrance}  \& {Sweet}}{{Reid}
  et~al.}{2008}]{Reid:2008}
{Reid} I.~N.,  {Cruz} K.~L.,  {Kirkpatrick} J.~D.,  {Allen} P.~R.,  {Mungall}
  F.,  {Liebert} J.,  {Lowrance} P.,   {Sweet} A.,  2008, \mn@doi [\aj]
  {10.1088/0004-6256/136/3/1290}, \href
  {http://adsabs.harvard.edu/abs/2008AJ....136.1290R} {136, 1290}

\bibitem[\protect\citeauthoryear{{Reiners} \& {Basri}}{{Reiners} \&
  {Basri}}{2008}]{Reiners:2008b}
{Reiners} A.,  {Basri} G.,  2008, \mn@doi [\apj] {10.1086/590073}, \href
  {http://adsabs.harvard.edu/abs/2008ApJ...684.1390R} {684, 1390}

\bibitem[\protect\citeauthoryear{{Reiners} \& {Basri}}{{Reiners} \&
  {Basri}}{2009}]{Reiners:2009b}
{Reiners} A.,  {Basri} G.,  2009, \mn@doi [\apj]
  {10.1088/0004-637X/705/2/1416}, \href
  {http://adsabs.harvard.edu/abs/2009ApJ...705.1416R} {705, 1416}

\bibitem[\protect\citeauthoryear{{Reiners} \& {Basri}}{{Reiners} \&
  {Basri}}{2010}]{Reiners:2010}
{Reiners} A.,  {Basri} G.,  2010, \mn@doi [\apj] {10.1088/0004-637X/710/2/924},
  \href {http://adsabs.harvard.edu/abs/2010ApJ...710..924R} {710, 924}

\bibitem[\protect\citeauthoryear{Richards, Waltman, Ghigo  \&
  Richards}{Richards et~al.}{2003}]{Richards:2003}
Richards M.~T.,  Waltman E.~B.,  Ghigo F.~D.,   Richards D. S.~P.,  2003,
  Astrophysical Journal Supplement Series, 147, 337

\bibitem[\protect\citeauthoryear{Robinson \& Melrose}{Robinson \&
  Melrose}{1984}]{Robinson:1984}
Robinson P.~A.,  Melrose D.,  1984, Australian Journal of Physics, 37, 675

\bibitem[\protect\citeauthoryear{{Rodr{\'{\i}}guez-Barrera}, {Helling}, {Stark}
   \& {Rice}}{{Rodr{\'{\i}}guez-Barrera} et~al.}{2015}]{Rodriguez-Barrera:2015}
{Rodr{\'{\i}}guez-Barrera} M.~I.,  {Helling} C.,  {Stark} C.~R.,   {Rice}
  A.~M.,  2015, \mn@doi [\mnras] {10.1093/mnras/stv2090}, \href
  {http://adsabs.harvard.edu/abs/2015MNRAS.454.3977R} {454, 3977}

\bibitem[\protect\citeauthoryear{{Route} \& {Wolszczan}}{{Route} \&
  {Wolszczan}}{2013}]{Route:2013}
{Route} M.,  {Wolszczan} A.,  2013, \mn@doi [\apj]
  {10.1088/0004-637X/773/1/18}, \href
  {http://adsabs.harvard.edu/abs/2013ApJ...773...18R} {773, 18}

\bibitem[\protect\citeauthoryear{{Sault}, {Teuben}  \& {Wright}}{{Sault}
  et~al.}{1995}]{Sault:1995}
{Sault} R.~J.,  {Teuben} P.~J.,   {Wright} M.~C.~H.,  1995, in {Shaw} R.~A.,
  {Payne} H.~E.,   {Hayes} J.~J.~E.,  eds,  Astronomical Society of the Pacific
  Conference Series Vol. 77, Astronomical Data Analysis Software and Systems
  IV. p.~433 (\mn@eprint {} {astro-ph/0612759})

\bibitem[\protect\citeauthoryear{{Scargle}}{{Scargle}}{1982}]{Scargle:1982}
{Scargle} J.~D.,  1982, \mn@doi [\apj] {10.1086/160554}, \href
  {http://adsabs.harvard.edu/abs/1982ApJ...263..835S} {263, 835}

\bibitem[\protect\citeauthoryear{{Schmidt}, {Cruz}, {Bongiorno}, {Liebert}  \&
  {Reid}}{{Schmidt} et~al.}{2007}]{Schmidt:2007}
{Schmidt} S.~J.,  {Cruz} K.~L.,  {Bongiorno} B.~J.,  {Liebert} J.,   {Reid}
  I.~N.,  2007, \mn@doi [\aj] {10.1086/512158}, \href
  {http://adsabs.harvard.edu/abs/2007AJ....133.2258S} {133, 2258}

\bibitem[\protect\citeauthoryear{{Schmidt}, {Hawley}, {West}, {Bochanski},
  {Davenport}, {Ge}  \& {Schneider}}{{Schmidt} et~al.}{2015}]{Schmidt:2015}
{Schmidt} S.~J.,  {Hawley} S.~L.,  {West} A.~A.,  {Bochanski} J.~J.,
  {Davenport} J.~R.~A.,  {Ge} J.,   {Schneider} D.~P.,  2015, \mn@doi [\aj]
  {10.1088/0004-6256/149/5/158}, \href
  {http://adsabs.harvard.edu/abs/2015AJ....149..158S} {149, 158}

\bibitem[\protect\citeauthoryear{{Stelzer} et~al.,}{{Stelzer}
  et~al.}{2012}]{Stelzer:2012}
{Stelzer} B.,  et~al., 2012, \mn@doi [\aap] {10.1051/0004-6361/201118097},
  \href {http://adsabs.harvard.edu/abs/2012A%26A...537A..94S} {537, A94}

\bibitem[\protect\citeauthoryear{{VanderPlas}, {Connolly}, {Ivezic}  \&
  {Gray}}{{VanderPlas} et~al.}{2012}]{VanderPlas:2012}
{VanderPlas} J.,  {Connolly} A.~J.,  {Ivezic} Z.,   {Gray} A.,  2012, in
  Proceedings of Conference on Intelligent Data Understanding (CIDU). pp 47--54
  (\mn@eprint {arXiv} {1411.5039}), \mn@doi{10.1109/CIDU.2012.6382200}

\bibitem[\protect\citeauthoryear{{West} et~al.,}{{West}
  et~al.}{2004}]{West:2004}
{West} A.~A.,  et~al., 2004, \mn@doi [\aj] {10.1086/421364}, \href
  {http://adsabs.harvard.edu/abs/2004AJ....128..426W} {128, 426}

\bibitem[\protect\citeauthoryear{{Williams} \& {Berger}}{{Williams} \&
  {Berger}}{2015}]{Williams:2015}
{Williams} P.~K.~G.,  {Berger} E.,  2015, \mn@doi [\apj]
  {10.1088/0004-637X/808/2/189}, \href
  {http://adsabs.harvard.edu/abs/2015ApJ...808..189W} {808, 189}

\bibitem[\protect\citeauthoryear{{Williams}, {Cook}  \& {Berger}}{{Williams}
  et~al.}{2014}]{Williams:2014}
{Williams} P.~K.~G.,  {Cook} B.~A.,   {Berger} E.,  2014, \mn@doi [\apj]
  {10.1088/0004-637X/785/1/9}, \href
  {http://adsabs.harvard.edu/abs/2014ApJ...785....9W} {785, 9}

\bibitem[\protect\citeauthoryear{{Wilson} et~al.,}{{Wilson}
  et~al.}{2011}]{Wilson:2011}
{Wilson} W.~E.,  et~al., 2011, \mn@doi [\mnras]
  {10.1111/j.1365-2966.2011.19054.x}, \href
  {http://adsabs.harvard.edu/abs/2011MNRAS.416..832W} {416, 832}

\makeatother
\end{thebibliography}

%\begin{thebibliography}{99}
%\end{thebibliography}

%%%%%%%%%%%%%%%%%%%%%%%%%%%%%%%%%%%%%%%%%%%%%%%%%%

%%%%%%%%%%%%%%%%% APPENDICES %%%%%%%%%%%%%%%%%%%%%

%\appendix


%t can be placed in an Appendix which appears after the list of references.

%%%%%%%%%%%%%%%%%%%%%%%%%%%%%%%%%%%%%%%%%%%%%%%%%%

% Don't change these lines
\bsp	% typesetting comment
\label{lastpage}
\end{document}